\documentclass[aps,prb,twocolumn,superscriptaddress,amsmath]{revtex4-1}

\usepackage[pdftex]{graphicx}
\usepackage{bm}

\begin{document}

\title{Rutgers relation for the analysis of superfluid density in superconductors}

\author{H. Kim}
\affiliation{The Ames Laboratory, Ames, IA 50011, USA}
\affiliation{Department of Physics \& Astronomy, Iowa State University, IA 50011, USA}

\author{V. G. Kogan}
\affiliation{The Ames Laboratory, Ames, IA 50011, USA}

\author{K. Cho}
\affiliation{The Ames Laboratory, Ames, IA 50011, USA}

\author{M. A. Tanatar}
\affiliation{The Ames Laboratory, Ames, IA 50011, USA}
\affiliation{Department of Physics \& Astronomy, Iowa State University, IA 50011, USA}

\author{R. Prozorov}
\email[Corresponding author: ]{prozorov@ameslab.gov}
\affiliation{The Ames Laboratory, Ames, IA 50011, USA}
\affiliation{Department of Physics \& Astronomy, Iowa State University, IA 50011, USA}

\date{28 May 2013}

\begin{abstract}
It is shown that the thermodynamic Rutgers relation for the second order phase transitions can be used for the analysis of the superfluid density data irrespective of complexities of the Fermi surface, structure of the superconducting gap, pairing strength, or scattering. The only limitation is that critical fluctuations should be weak, so that the mean field theory of the second order phase transitions is applicable. By using the Rutgers relation, the zero temperature value of the London penetration depth, $\lambda(0)$, is related to the specific heat jump $\Delta C$ and the slope of upper critical field $dH_{c2}/dT$ at the transition temperature $T_c$, provided the data on $\Delta\lambda=\lambda(T)-\lambda(0)$ are available in a broad temperature domain. We then provide a new way of determination of $\lambda(0)$, the quantity difficult to determine within many techniques.
\end{abstract}



\maketitle

\section{ Introduction }

The London penetration depth  $\lambda$ is one of the most important characteristic length scales of superconductors. The temperature dependent $\lambda(T)$ is subject of many studies, as it provides information on the symmetry of the order parameter.\cite{Prozorov2006,Prozorov2011} It is used to calculate the superfluid density, $\rho(T) \equiv \lambda^2(0)/\lambda^2(T)$, a quantity that can be directly compared with theory. Determination of $\lambda(0)$ is critical because the shape of $\rho(T)$ extracted from the data on $\Delta\lambda(T) $ depends sensitively on the value of $\lambda(0)$ adopted, and a wrong $\lambda(0)$ could lead to incorrect conclusions on the superconducting order parameter.

Tunnel diode resonator (TDR) provides perhaps the most precise measurements of the variation of $\lambda$ with temperature, $\Delta\lambda=\lambda(T)-\lambda(0)$. With additional sample manipulation by coating it with lower-$T_c$ superconductor, the absolute value of $\lambda (0)$ can be determined as well, but with much lower precision compared to $\Delta \lambda$.\cite{Prozorov2000coating}
Other techniques which are used to measure $\lambda(0)$ include muon spin rotation ($\mu$SR),\cite{Sonier2007} infrared spectroscopy,\cite{Basov2005} and microwave cavity perturbation technique,\cite{Kamal1998} and local probes.\cite{Lippman2012,Luan2010} However, each of these techniques has its own limitations. $\mu$SR measures averaged $\lambda(T,H)$ in the mixed state and extrapolated to $H=0$ value is used to estimate $\lambda(0)$. In infrared spectroscopy, $\lambda(0)$ is deduced from the measured plasma frequency, which is not a precisely determined quantity.\cite{Basov2005} Local probes, such as scanning SQUID \cite{Lippman2012} and MFM \cite{Luan2010} magnetometry, infer $\lambda (0)$ from the analysis of magnetic interactions between a relevant probe and a magnetic moment induced in a superconductor.\cite{Kirtley2010}

In this paper we show that the thermodynamic relation between the specific heat jump, $\Delta C$, and the slope of thermodynamic critical field, $\partial H_c/\partial T$, at the superconducting transition temperature, $T_c$, first proposed by Rutgers \cite{Rutgers1934}, can be rewritten in terms of measurable quantities, - the slopes of the upper critical field and of the superfluid density in addition to specific heat jump. As a general thermodynamic relation valid at the 2nd order transition (excluding critical fluctuation region), it is applicable for any superconductor irrespective of the pairing symmetry, scattering or multiband nature of superconductivity, as we verified on several well-known systems. Such general applicability of the Rutgers relation offers a method of estimating $\lambda(0)$ if $\Delta C$ and $dH_{c2}/dT$ at $T_c$ are known. This idea is checked on Nb and MgB$_2$ and applied to several unconventional superconductors. In all cases we use $\Delta\lambda (T)$ measured by the TDR technique and the literature data for the other two quantities except for YBa$_2$Cu$_3$O$_{1-\delta}$ where its $\rho(T)$ was taken from Ref.\onlinecite{Kamal1998}. In all studied cases, the method works well and determined values of $\lambda(0)$ are in agreement with the literature.

\subsection{Thermodynamic Rutgers relation}

 The specific heat jump at $T_c$ in materials where the critical fluctuations are weak is expressed through the free energy difference $F_n-F_s=H_c^2/8\pi$:\cite{Rutgers1934,Lifshitz1984}    
%
 \begin{eqnarray}
  \Delta C = T_c\frac{\partial^2}{\partial T^2} \frac{H_c^2}{8\pi }\Big|_{T_c}=\frac{ T_c}{4\pi}\left( \frac{\partial H_c}{\partial T}\right)^2_{T_c}  
  \label{Rutgers}
   \end{eqnarray}
Here, $C$ is measured in erg/cm$^3$K and $T$ in K. 
 Within the mean-field Ginzburg-Landau (GL) theory, near $T_c$, the thermodynamic critical field  $H_c=\phi_0/2\sqrt{2}\pi\xi\lambda$ with    
 \begin{eqnarray}
( \xi,\lambda) =\frac{( \xi_{GL},\lambda_{GL})}{\sqrt{1-t}} \,,\qquad t=\frac{T}{T_c}\,.
\label{GLs}
   \end{eqnarray}
  Here $\xi$ and $\lambda$ are the coherence length and the penetration depth, and  the constants $ \xi_{GL},\lambda_{GL} $ are of the same order  but not the same as the zero$-T$ values $ \xi(0)$ and $\lambda(0) $. 
 Hence we have:
 \begin{eqnarray}
  \Delta C =\frac{ \phi_0^2}{32\pi^3 \xi_{GL}^2\lambda_{GL}^2T_c} \,,
  \label{DC}
   \end{eqnarray}
   where  $\xi_{GL}$ is related to the   slope  of $H_{c2}(T)$  at $T_c$:
 \begin{eqnarray}
T_c \frac{\partial H_{c2}}{\partial T}\Big |_{T_c} =\frac{\partial H_{c2}}{\partial t}\Big |_{t=1}=-\frac{\phi_0}{ 2\pi  \xi_{GL}^2 } \,.
   \end{eqnarray}
  It is  common  to introduce the dimensionless superfluid density $\rho=\lambda^2(0)/\lambda^2$ with the slope at $T_c$ given by
  \begin{eqnarray}
  T_c \frac{\partial  \rho}{\partial T}\Big |_{T_c}  =\frac{\partial  \rho}{\partial t}\Big |_{t=1}  =-\frac{ \lambda ^2(0)}{\lambda_{GL}^2 }\,.
   \end{eqnarray}
We then obtain:
 \begin{eqnarray}
  \Delta C =\frac{ \phi_0 }{16\pi^2  \lambda ^2(0)T_c   } \, \left(H_{c2}^\prime \, \rho^\prime  \right)_{t=1} 
    \label{Rut}
   \end{eqnarray}
\noindent where the primes denote derivatives with respect to $t$. 
 
 It should be stressed that being a thermodynamic relation that holds at a 2nd order phase transition, applicability of Rutgers formula is restricted only by possible presence of critical fluctuations. In particular, it can be applied for zero-field phase transition in materials with anisotropic order parameters and Fermi surfaces, multi-band etc, which makes it a valuable tool in studying great majority of new materials.
 
   For anisotropic materials, Eq.\,(\ref{Rutgers}) is, of course, valid since the condensation energy and $H_c$ do not depend on direction. However, already in Eq.\,(\ref{DC}) the field direction should be specified.    In the following we discuss situations with ${\bm H}$ parallel to the $c$ axis of uniaxial crystals. Hence, $H_{c2}, \rho$, and $\lambda(0)$ in Eq.\,(\ref{Rut}) should have subscripts $ab$; we omit them for brevity. A general case of anisotropic material with arbitrary field orientation requires separate analysis.



\section{ Determination of $\lambda(0)$ } 

The full superfluid density needed for the analysis of the experimental data and comparison with theoretical calculations depends on the choice of $\lambda(0)$:
\begin{equation}
\label{rho}
\rho(t)=\frac {\lambda^2(0)}{[\lambda(0)+\Delta\lambda (t)]^2}  \,.
\end{equation}
%
\begin{figure}
\includegraphics[width=1.0\linewidth]{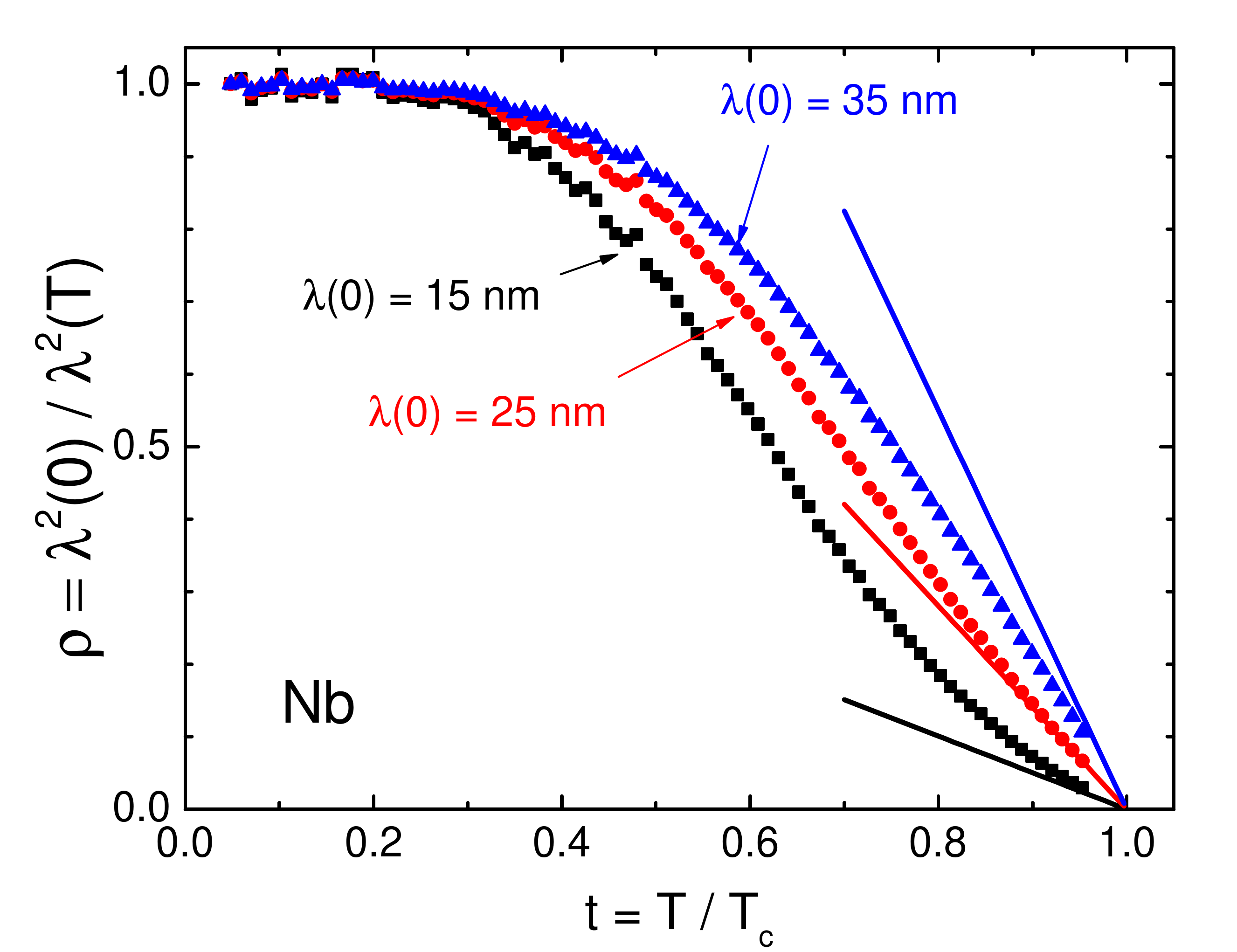}%
\caption{\label{fig1} (Color online) Superfluid density $\rho (t)$ calculated from Eq.\,(\ref{rho}) using the TDR data on $\Delta\lambda (T)$ and assuming  $\lambda(0)=$ 15, 25, and 35 nm. Straight lines have the slope $\rho'$ estimated from Eq.~(\ref{Rut}) for each $\lambda(0)$. }
\label{fig1}
\end{figure}
Figure \ref{fig1} shows an example of this dependence of $\rho(t)$ on $\lambda(0)$ for Nb. Symbols represent $\rho(t)$ calculated from measured $\Delta\lambda(t)$ with $\lambda(0)$ chosen as 15,  25 and 35 nm. Clearly, the calculated $\rho(t)$ is sensitive to the choice of $\lambda(0)$. The straight solid lines have the slope $\rho'(1)$ calculated by using Eq.~(\ref{Rut}) for each $\lambda(0)$. We used $\Delta C=137.2$ mJ/mol-K  $ = 126450$ erg/cm$^3$K (Ref. \onlinecite{Weber1991}) since in the formulas used here the specific heat is per unit volume. \footnote{To convert $\Delta C$ which is commonly reported in mJ/mol-K into erg/cm$^3$K, one needs to calculate the mass density which requires crystallographic information. For niobium we use parameters found in Ref.~\onlinecite{Ito2006}. Crystal structure of elemental niobium belongs to the space group Im-3m (no. 229) with lattice parameters $a = b = c = 0.3303$ nm, and corresponding volume is $V=0.036$ nm$^3$. There are two molecular units per the volume ($Z=2$). Using these values the converted $\Delta C=137.2$ mJ/mol-K (Ref. \onlinecite{Weber1991}) $=126450$ erg/cm$^3$K.} Using $H_{c2}'|_{T_c}=440$ Oe/K (Ref. \onlinecite{Williamson1970}), we obtain $-\rho'(1)=$ 0.49, 1.4, and 2.7 for 15, 25, and 35 nm, respectively. While the choice of $\lambda(0)=25$ nm shows reasonable agreement, for the choices of 15 nm and 35 nm the slopes calculated using the data and Eq.~(\ref{rho}) determined by Eq.~(\ref{Rut}) under- and over-estimates, respectively. Note that with $\lambda(0)=15$ nm, the temperature dependence of $\rho$ is pronouncedly concave near $t=1$, and also $-\rho'(1)$ is smaller than one. The idea of our method is to utilize the Rutgers relation (\ref{Rut}) and  choose such a $\lambda(0)$ that would not contradict the thermodynamics near $T_c$. 
 
To this end we rewrite Eq.\,(\ref{Rut}) in the form:
\begin{equation} 
\frac{ \rho'(1)}{\lambda^2(0)}=\frac{16\pi^2 T_c \Delta C}{ \phi_0  H^\prime_{c2}(1)}\,.
\label{Ruta}
\end{equation}
The right-hand side here is determined from independent measurements of $\Delta C$ and $H_{c2}$. Thus, by taking a few test values of $\lambda(0)$, calculating  $\rho(t)$ and its slope at $t=1$, we can decide which $\lambda(0)$ and $\rho(t,\lambda(0))$ obey the Rutgers relation. 

\begin{figure*}[htb]
\includegraphics[width=0.8\linewidth]{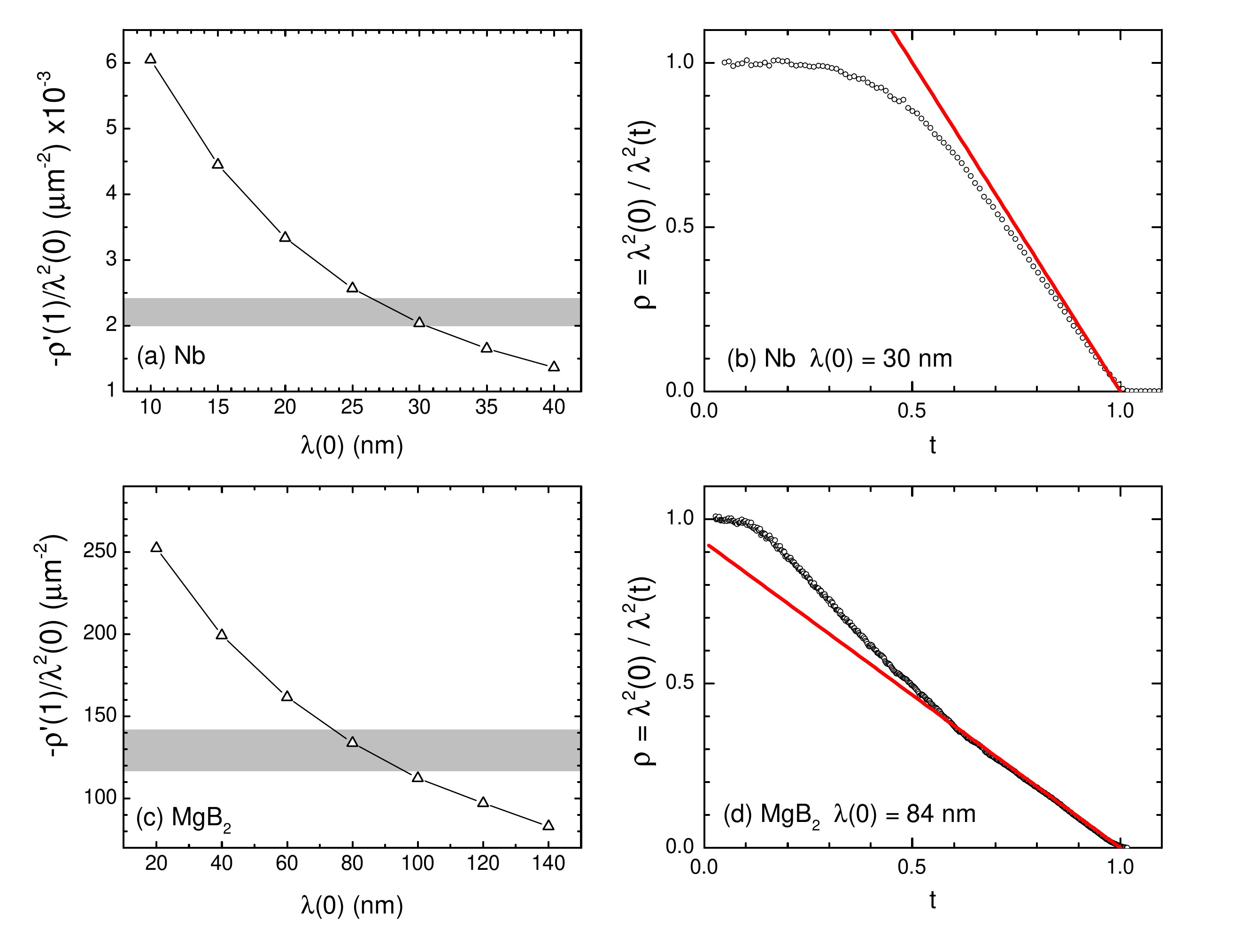}%
\caption{Top row - Nb, Bottom row - MgB$_2$. Left column: variation of $|\rho'|/\lambda^2(0)$ as a function of $\lambda(0)$ for Nb and MgB$_2$. Shaded horizontal bands are the estimated values of the right-hand side of Eq.\,(\ref{Ruta}) with  literature values of $\Delta C$ and $H_{c2}^\prime(1)$ including experimental uncertainties. Right column: superfluid density for the best value of $\lambda(0)$ that satisfies the Rutgers relation, Eq.\,(\ref{Ruta}).}
\label{fig2}
\end{figure*}

We first apply this method to two well-studied superconductors - conventional Nb and two-band MgB$_2$. For Nb, we obtain $|\rho'|/\lambda^2(0)\approx 2240$ $\mu$m$^{-2}$ using the same thermodynamic quantities as for Fig.~\ref{fig1}. \cite{Weber1991,Williamson1970} We now take a set of values for $\lambda(0)$ shown in top left panel of Fig.\,\ref{fig2} and plot $|\rho^\prime|/\lambda^2(0)$ vs $\lambda(0)$. The value of $\lambda(0)= 28\pm 2\,$nm satisfying the Rutgers relation is obtained from the intersection of the calculated curve with the value expected from Eq.\,(\ref{Ruta}) (shown by a gray band that takes into account experimental uncertainties in determining $\Delta C$ and $H'_{c2}$). It is consistent with the literature values varying between 26 and 39 nm.\cite{Weber1991,Maxfield1965} The final calculated superfluid density with the choice of $\lambda(0)=30$ nm is shown in Fig.~\ref{fig2}(b). The solid line is determined with the calculated slope $|\rho'(1)|=2$, as predicted for isotropic s-wave superconductors (see, Appendix).

In addition to the aforementioned uncertainties, determination of the experimental $|\rho'(1)|$ is not trivial even if the quality of measurement is excellent since $\rho(t)$ near $t=1$ is often significantly curved due to several experimental artifacts, most importantly due to the influence of the normal skin effect near $T_c$, which is more pronounced for higher frequency measurements on highly conducting materials. TDR technique uses typically $\sim 10$ MHz, so this effect is weak in most of the materials concerned. Analyzing the data for different superconductors, we have found that the data in the regime between $t=0.8$ and 0.95 works well for the determination of $\rho'(1)$. The experimental $|\rho'(1)|$ in this work is determined from the best linear fit of $\rho(t)$ data in this range. 

The same procedure can be employed for a well known multi gap superconductor MgB$_2$ (shown in the bottom row of Fig.~\ref{fig2}), where $|\rho'|/\lambda^2(0)$ is estimated to be $130 \pm 12$ $\mu$m$^{-2}$ by using $\Delta C=133$ mJ/mol-K (Ref. \onlinecite{Bouquet2001}), $|H'_{c2}(1)| = 0.45$ T/K (Ref. \onlinecite{Budko2001}) within $\pm$5\% error. The determined $\lambda(0)= 84\pm 10\,$nm is in good agreement with 100 nm estimated by $\mu$SR technique.\cite{Niedermayer2002,Ohishi2003} For $\lambda(0)= 84$ nm, the calculated slope $|\rho'(1)|=0.91$ agrees with the expected theoretical value of 0.92. (appendix).

The method described has also been used  for SrPd$_2$Ge$_2$ for which $\lambda(0)$ was not clear. By using the determined $\lambda(0)$ we have shown that SrPd$_2$Ge$_2$ is a single-gap s-wave superconductor.\cite{Kim2013}




\subsection{Unconventional superconductors}
\begin{table*}[thb] 
\caption{$V_c$ is the volume of the unit cell. $\Delta C$ is the specific heat jump at $T_c$ in mJ/mol-K. $dH_{c2}/dT$ is slope of $H_{c2}$ at $T_c$. $\rho'_{Rut}=(d\rho/dt)_{Rut}$ is the calculated slope using Eq.\,(\ref{Rut}) where $t=T/T_c$. $\rho'_{exp}$ is an experimental slope with given $\lambda(0)$.}
\label{tab}
\begin{ruledtabular}
\begin{tabular}{l|ccccccc}
compound & $V_{c}$ \,(\text{\AA}$^3$)& $T_c$ (K) & $\Delta C/T_c$ (mJ/mol-K$^2$) &$|dH_{c2}/ dT| _{T_c}$\,(T/K) & $\lambda(0)$ (nm)  & $-\rho'_{Rut}$ & $-\rho'_{exp}$\\
\hline \hline

Nb & 35.937 [\onlinecite{Ito2006}]& 9.3 [\onlinecite{Weber1991}] &  14.8 [\onlinecite{Weber1991}] &  0.044 [\onlinecite{Williamson1970}] & 30\footnote{determined in this work.} & 2.0  & 1.8\\

MgB$_{2}$ & $29.064$ [\onlinecite{Nagamatsu2001}]& 39 [\onlinecite{Bouquet2001}]  & 3.4 [\onlinecite{Bouquet2001}] & $0.45$ [\onlinecite{Budko2001}] & 84\footnote{determined in this work.} &0.91  & $0.83$\\ \hline

LiFeAs & $90.252$ [\onlinecite{Tapp2008}] & 15.4 [\onlinecite{Wei2010}] & 20 [\onlinecite{Wei2010}] & 3.46 [\onlinecite{Cho2011}] & 200 [\onlinecite{Pratt2009}] & 1.2 & 1.1 \\

FeTe$_{0.58}$Se$_{0.42}$ & 87.084 [\onlinecite{Johnston2010}]& 14 [\onlinecite{Braithwaite2010}]  & 20 [\onlinecite{Braithwaite2010}] &13 [\onlinecite{Klein2010}]& 500\footnote{an average value over 430-560 nm (Ref. \onlinecite{Klein2010,Biswas2010,Kim2010})} & 1.4 & 1.5\\

YBa$_2$Cu$_3$O$_{1-\delta}$ & 173.57 [\onlinecite{Poole2010}]& 23 [\onlinecite{JKim2012}] & 61 [\onlinecite{Wang2001}]& 1.9 [\onlinecite{Welp1989}] & 120 [\onlinecite{Prozorov2000coating}] & 3.0 & 2.15 - 4.98 [\onlinecite{Kamal1998}] \\

MgCNi$_3$ & 54.496 [\onlinecite{Shan2003}]& 7 [\onlinecite{Shan2003}]& 129 [\onlinecite{Shan2003}]& 2.6 [\onlinecite{Poole2010}]& 232 [\onlinecite{MacDougall2006}]&1.8 & 2.0 \\
\end{tabular}
\end{ruledtabular}
\end{table*}
%
Here we examine the validity of our approach for some superconductors for which the necessary experimental quantities have been reported in the literature. Where possible, we use $H_{c2}(T)$ determined from the specific heat jump, because resistive and magnetic measurements may actually determine the irreversibility field, which may differ substantially from the thermodynamic $H_{c2}$.\cite{Carrington1996}

We have selected LiFeAs, FeTe$_{0.58}$Se$_{0.42}$, YBa$_2$Cu$_3$O$_{1-\delta}$ and MgCNi$_3$ representing stoichiometric pnictide, charchogenide, d-wave high-$T_c$ cuprate and close to magnetic instability s-wave superconductor, respectively. $\Delta C$, $dH_{c2}/dT$, and $\lambda(0)$ for the selected compounds have been measured by various techniques by different groups. The superfluid density was calculated from the penetration depth measured by using a TDR technique at Ames Laboratory, except for YBCO for which anisotropic superfluid density was determined by microwave cavity perturbation technique.\cite{Kamal1998} Thermodynamic parameters are discussed in the number of papers.\cite{Johnston2010,Canfield2010,Stewart2011} In-depth discussion of the specific heat is given in Refs.~\onlinecite{Stewart2011,JKim2012}. Table \ref{tab} summarizes parameters used in the calculations. 

\begin{figure*}[tbh]
\includegraphics[width=0.8\linewidth]{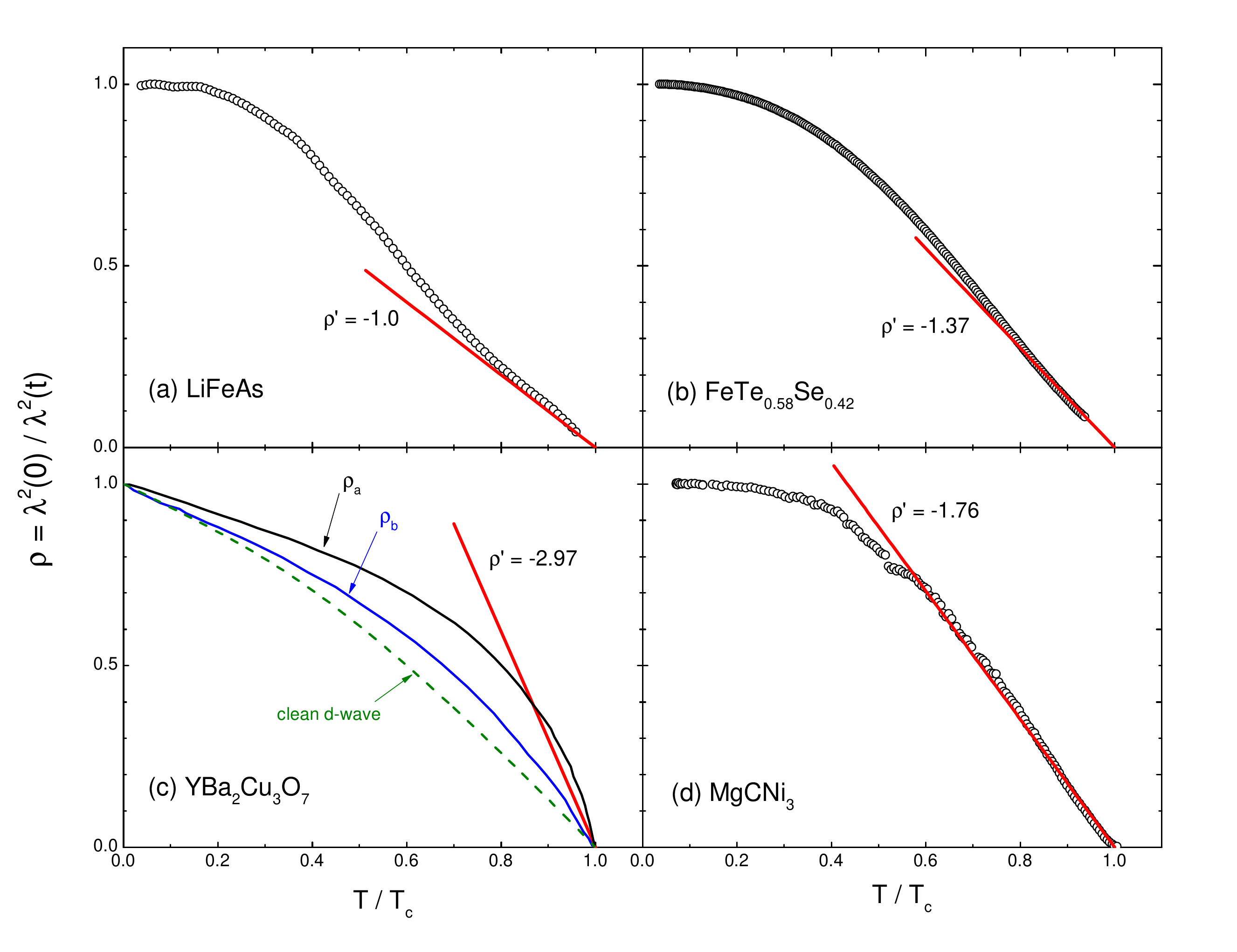}%
\caption{Experimental superfluid density $\rho=\lambda^2(0)/\lambda^2(T)$ in LiFeAs, FeTe$_{0.58}$Se$_{0.42}$, YBa$_2$Cu$_3$O$_{1-\delta}$, and MgCNi$_3$ with $\lambda(0)=$ 500, 200, 120, and 232 nm, respectively. The straight lines in each panel were estimated with the Rutgers formula. Parameters used for the calculation are summarized in Table \ref{tab}.}
\label{fig}
\end{figure*}

Figure \ref{fig} shows experimental superfluid density in LiFeAs, FeTe$_{0.58}$Se$_{0.42}$, YBa$_2$Cu$_3$O$_{1-\delta}$ and MgCNi$_3$ with $\lambda(0)=500$, 200, 120, and 232 nm, respectively. The agreement between $\rho^\prime_{Rut}$ calculated with the Rutgers relation and $\rho^\prime_{exp}$ extracted from the data on $\Delta\lambda(t)$, given possible uncertainties in the input experimental parameters, is rather remarkable. 




\section{Summary}

In conclusion, we have shown that the thermodynamic relation based on Rutgers formula can be used for the analysis of the superfluid density. Based on this relation, a method to estimate $\lambda(0)$ is developed. As a test, it was successfully applied to reproduce known $\lambda(0)$ in Nb and MgB$_2$. We used this relation to verify reported literature values of $\lambda(0)$ for several unconventional superconductors of different band structure, gap anisotropy, and pairing symmetry. 

\section{ACKNOWLEDGMENTS}
We thank A. Chubukov for useful discussions. The work was supported by the U.S. Department of Energy, Office of Basic Energy Sciences, Division of Materials Sciences and Engineering under contract No. DE-AC02-07CH11358.

\appendix

\section{Theoretical results relevant for the analysis of the superfluid density}

 \subsection{Penetration depth in anisotropic materials}
 
It is known\cite{Parks1969} that in isotropic materials, 
\begin{equation}
\rho^\prime(1)=  \lambda ^2(0)/\lambda_{GL}^2=-2\,.
\label{rho'is}
\end{equation}
It is easy to reproduce this result for the  free electron model of the normal state; it is shown below, however, that this value holds for any Fermi surface provided the order parameter is isotropic. 
 
Here, we are interested in relating  $\lambda(0)$ and $\lambda_{GL}$,  the $T$ independent part of $\lambda$ near $T_c$,    for anisotropic Fermi surfaces and order parameters. 
We start with a known relation, 
\begin{equation}
(\lambda^2)_{ik}^{-1}= \frac{16\pi^2 e^2N(0) T}{
c^2} \sum_\omega \left\langle \frac{\Delta^2v_iv_k}{\beta^3}\right\rangle \,,  
\label{lambda-tensor}
\end{equation}
\noindent which holds at any temperature for {\it clean} materials with arbitrary Fermi
surface and order parameter anisotropies.\cite{Kogan2002,Prozorov2011} Here, $N(0)$ is the density of states at the Fermi level per spin, $\beta^2=\Delta^2+\hbar^2\omega^2$ with $\hbar \omega=\pi T(2n+1)$, $\Delta({\bm k}_F,T)=\Psi(T)\Omega({\bm k}_F) $ is the zero-field order parameter which in general depends on the position ${\bm k}_F$ on the Fermi surface, and $\langle...\rangle$ stand for averaging over the Fermi surface.  The  function $\Omega({\bm  k}_F)$  which describes the variation of
$\Delta$ along the Fermi surface, is normalized:
$\left\langle \Omega^2 \right\rangle=1$.

Eq.\,(\ref{lambda-tensor}) is obtained within the model of factorizable effective coupling $ V({\bm k},{\bm k}^{\prime })=V_0 \,\Omega({\bm k})\,\Omega({\bm k}^{\prime })$.\cite{Markowitz1963} The self-consistency equation of the weak coupling theory takes the form:
\begin{equation}
\Psi( {\bm  r},T)=2\pi T N(0)V_0 \sum_{\omega >0}^{\omega_D} \Big\langle
\Omega({\bm k} ) f({\bm k} ,{\bm  r},\omega)\Big\rangle \,,
\label{gap}
\end{equation}
where $f$ is the Eilenberger Green's function which, for the uniform current-free state, reads: $f=\Delta/\beta= \Psi\Omega/\beta$. The order parameter near $T_c$ is now readily obtained:
        \begin{equation}
\Psi^2=  \frac{8\pi^2T_c^2(1-t)} {7\zeta(3)\langle 
\Omega^4 \rangle} \,,
\label{Psi(Tc)}
\end{equation}
which  reduces to the isotropic BCS form for $\Omega=1$. 
We substitute this in Eq.\,(\ref{lambda-tensor}) to obtain near $T_c$:
\begin{equation}
(\lambda^2)_{ik}^{-1}= \frac{16\pi e^2N(0) \langle \Omega^2v_iv_k  \rangle}{c^2 \langle \Omega^4  \rangle}\,(1-t) \,,
       \label{ Tc-GL}
\end{equation}
\noindent from which the constants $\lambda_{GL}$ for any direction  readily follow. 


As $T\to 0$, the sum over the Matsubara frequencies in Eq.\,(\ref{lambda-tensor}) can be replaced with an integral according to $2\pi
T \sum_{\omega}\to \int_0^\infty d(\hbar\omega)$:
\begin{equation}
(\lambda^2)_{ik}^{-1} (0)= \frac{8\pi  e^2 N(0)}{c^2}\, \Big\langle
      v_iv_k \Big\rangle \,.  \label{T=0}
\end{equation}
For free electrons, this reduces to the London value $\lambda^2=mc^2/4\pi e^2 n$ where $n=2m N(0)v^2/3$ is the electron density.

Hence, we get for the slope of the in-plane superfluid density:
 \begin{equation}
\rho_{ab}^\prime(1)=-\frac{\lambda^{2}_{ab} (0)}{\lambda^{2}_{GL,ab} }=-2 \frac{  \langle \Omega^2 v^2_a  \rangle} { \langle v_a^2   \rangle \langle \Omega^4  \rangle} \,. 
       \label{ratios}
\end{equation}
Similarly, one can define $\rho_c^\prime(1)$ for which $v_a$ should be replaced with $v_c$ in Eq.\,(\ref{ratios}). In particular, we have:
 \begin{equation}
\frac{\rho_c^\prime(1)}{\rho_{ab}^\prime(1)}=  \frac{  \langle  v^2_a  \rangle} { \langle v_c^2   \rangle }\, \frac{  \langle \Omega^2 v^2_c  \rangle} {  \langle \Omega^2 v^2_a  \rangle } =\frac{\gamma_\lambda^2(0)}{\gamma_\lambda^2(T_c)}\,. 
       \label{ratios1}
\end{equation}
E.g., for MgB$_2$ with $\gamma_\lambda (0)\approx 1$, $\gamma_\lambda (T_c)\approx 2.6$, we estimate $\rho_c^\prime(1) \approx 0.15\, \rho_{ab}^\prime(1)$.

It is instructive to note that $\rho^\prime(1)$ reduces to the  isotropic value of $-2$  for  {\it any} Fermi surface provided the   order parameter is constant, $\Omega=1$.

\subsection{MgB$_2$}

  Consider a simple two-band model with the gap anisotropy given by
\begin{equation}
\Omega ({\textbf k})= \Omega_{1,2}\,,\quad {\textbf k}\in   F_{1,2} \,, 
 \label{e50} 
\end{equation}
where $F_1,F_2$ are two sheets of the Fermi surface.  $ \Omega_{1,2}$ are assumed constants, in other words, we model  MgB$_2$ as having two different s-wave gaps.  
  The normalization $\langle\Omega^2\rangle=1$ then gives: 
\begin{equation}
\Omega_1^2 \nu_1+\Omega_2^2\nu_2=1\,,\quad \nu_1+\nu_2=1\,,\label{norm1}
\end{equation}
where $\nu_{1,2}= N_{1,2}/N(0)$ are the relative densities of states.
 
Based on the band structure calculations,\cite{Belashchenko2001,Choi2002}   
$\nu_1$ and $\nu_2$ of our model are $\approx\,$0.56 and 0.44.
The ratio $\Delta_2/\Delta_1=\Omega_2/\Omega_1\approx
3$.
Then, the normalization (\ref{norm1})  yields
$\Omega_1=0.47$  and  $\Omega_2=1.41$. 

Further, we use the averages over
separate Fermi sheets calculated in Ref.\,\onlinecite{Belashchenko2001}: 
$\langle v_a^2\rangle_1=33.2$,  
$\langle v_a^2\rangle_2=23$\,cm$^2$/s$^2$.
With this input, we  estimate
 \begin{equation}
\rho_{ab}^\prime(1)= - 0.92 \,. 
       \label{ratiosMgB2}
\end{equation}
It should be noted that this number is sensitive to a number of input parameters. The procedure described above, see Fig.~\ref{fig2}, gives $\rho_{ab}^\prime(1)\approx - 0.91$.


Since only even powers of $\Omega$ enter Eq.\,(\ref{ratios}), the same analysis of the slope $\rho^\prime(1)$ can, in fact, be exercised for materials modeled by two bands with the $\pm s $ symmetry of the order parameter, for which $\Omega$'s have opposite signs.  
If the bands relative densities of state $\nu_{1,2}$ and the averages $\langle v_a^2\rangle_{1,2}$ are comparable to each other and similar to those of MgB$_2$, we expect a similar  $|\rho^\prime(1)|\approx 1$ for clean crystals. 

\subsection{d-wave}

  It can be shown that $\Omega=\sqrt{2}\cos 2\phi$   for closed Fermi surfaces as rotational ellipsoids  (in particular, spheres) or open ones as rotational hyperboloids  (in particular, cylinders).\cite{Kogan2012}  A straightforward algebra gives:
\begin{equation}
\rho^\prime_{ab}(1) =-4/3\,.
       \label{d-ratios}
\end{equation}

\subsection{Scattering}

In the  limit of a strong non-magnetic scattering for an arbitrary Fermi surface but a constant s-wave order parameter we have, see, e.g, Ref.\,\onlinecite{Prozorov2011}:
\begin{equation}
(\lambda^2)_{ik}^{-1}= \frac{8\pi^2 e^2 N(0)
\langle v_iv_k\rangle\tau}{
c^2\hbar}\,  \Delta  \tanh\frac{\Delta }{ 2T}  \,.  
\label{lambda-dirty}
\end{equation}
Here $\tau$ is the average scattering time. 
It is worth noting that the dirty limit does not make much sense for anisotropic gaps because $T_c$ is suppressed even by non-magnetic scattering  in the limit $\tau\to 0$.
At $T=0$, we have
\begin{equation}
(\lambda^2)_{ik}^{-1}(0)= \frac{8\pi^2 e^2 N(0)
\langle v_iv_k\rangle\tau}{
c^2\hbar}\,  \Delta (0)\,,
\label{T=0dirty}
\end{equation}
whereas near $T_c$
\begin{equation}
(\lambda^2)_{ik}^{-1} = \frac{8\pi^2 e^2 N(0)
\langle v_iv_k\rangle\tau}{
c^2\hbar}\, \frac{ \Delta^2  }{2 T_c}\,,
\label{Tc=dirty}
\end{equation}
Since for non-magnetic scattering, $T_c$ and $\Delta(T)$ are the same as in the clean case, in particular $\Delta =8\pi^2T_c^2(1-t)/7\zeta(3)$, we obtain
\begin{equation}
\rho^\prime (1)= -  \frac{4\pi^2 T_c}{\Delta(0)} =- \frac{4\pi  e^\gamma}{7\zeta(3)}=-2.66\,.
\label{rhjo'dirty}
\end{equation}
We thus conclude that scattering causes the slope $\rho^\prime (1)$ to increase.

Evaluation of scattering effects on the slope $\rho^\prime$ near $T_c$ for anisotropic gaps and Fermi surfaces are more involved because both $T_c$ and $\Delta$ are affected even by non-magnetic scattering. The case of a strong pair-breaking is an exception: $\lambda^{-2}=\lambda^{-2}_0(1-t^2)$ that immediately gives $\rho^\prime(1)=-2$.

\end{document}